\documentclass[11pt]{article}
\usepackage{latexsym}
\usepackage{amssymb}
\usepackage{amsmath}
\usepackage{amscd}
\usepackage{amsthm}
\usepackage[left=2cm,top=2.5cm,right=2.5cm,bottom=1.5cm]{geometry}

\usepackage[dvips]{graphicx}
\usepackage{hyperref}

\begin{document}
\begin{center}
\large{\bf{$\Lambda$CDM-type cosmological model and observational constraints}}\\
\vspace{6mm}
\normalsize{G. K. Goswami$^1$, Anil Kumar Yadav$^2$ and Mandwi Mishra$^3$}\\
\vspace{6mm}
\normalsize{$^1$ Kalyan Post-Graduate College, Bhilai, C.G., India}\\
\normalsize{Email: gk.goswam9i@gmail.com}\\
\vspace{2mm}
\normalsize{$^2$ Anand Engineering College, Keetham, Agra - 282007, India}\\
\normalsize{Email: abanilyadav@yahoo.co.in}\\
\vspace{2mm}
\normalsize{$^3$ Shri Shankaracharya Engineering College, Bhilai, C. G., India}
\end{center}
\vspace{2mm}
\begin{abstract}
In the present work, we have searched the existence of $\Lambda$CDM-type cosmological model in anisotropic 
Heckmann-Schucking space-time. The matter source that is responsible for the present acceleration 
of the universe consist of cosmic fluid with $p = \omega\rho$, where $\omega$ is the equation of 
state parameter. The Einstein's field equations have been solved explicitly under some specific choice 
of parameters that isotropizes the model under consideration. It has been found that the derived model 
is in good agreement with recent SN Ia observations. Some physical aspects of the model has been discussed 
in detail.\\


\end{abstract}

\section{Introduction}
The SN Ia observations \cite{Riess1998,Perlmutter1999} suggest that the observable 
universe is undergoing an accelerated expansion. This remarkable discovery 
stands a major break through of the observational cosmology and indicates the presence 
of unknown fluid - dark energy (DE) that opposes the self attraction of the matter. 
This acceleration is realized with positive energy density and negative pressure. 
So, it violate the strong energy condition (SEC). The authors of ref. \cite{caldwell2006} confirmed that 
the violation of SEC gives a reverse gravitational effect that provides an elegant description 
of transition of universe from deceleration to cosmic acceleration. 
The cosmological constant cold dark matter ($\Lambda$CDM) cosmological model is the simplest 
model of universe that describes the present acceleration of universe and fits with the present 
day cosmological data \cite{gron2007}. It is based on the Einstein's theory of general relativity 
with a spatially flat, isotropic and homogeneous space-time. The observed acceleration of universe 
has been explained by introducing a positive cosmological constant $\Lambda$ which is mathematically 
equivalent to vacuum energy with equation of state (EOS) parameter set equal to $-1$. It suffers from 
two problems on theoretical front, concerning the cosmological constant $(\Lambda)$. These problems are 
known as fine tunning and cosmic coincidence problems \cite{carroll1992,copeland2006}. In the 
contemporary cosmology, the source that derives the present acceleration of universe is still mystery and 
is discussed under the generic name DE. In the literature, the simplest candidate of dark energy is a positive 
$\Lambda$ besides some scalar field DE models, namely the phantom, quintessence and 
k-essence \cite{copeland2006,alam2003}. In the physical cosmology, the dynamical form of DE with an 
effective equation of state (EOS), $\omega < -\frac{1}{3}$, were proposed instead of constant 
vacuum energy. The current cosmological data from large scale-structure \cite{komastu2009}, Supernovae 
Legacy survey, Gold Sample of Hubble Space Telescope \cite{Riess2004,Astier2006} do not support the 
possibility of $\omega << -1$. However, $\omega = -1$ is a favorable candidate for DE that crossing the 
phantom divide line (PDL). Setare and Saridakis \cite{setare2008,setare2009} have studied 
the quintom model that described the 
nature of DE with $\omega$ across -1 and give the concrete theoretical justification for existence of quintom model\\

We notice that after publication of WMAP data, today there is considerable evidence in support of 
anisotropic model of universe. On the theoretical front, Misner \cite{misner1968} has investigated 
an anisotropic phase of universe, which turns into isotropic one. The authors of ref. \cite{koivisto2008,mota2008}
have investigated the accelerating model of universe with anisotropic EOS parameter and have also shown that 
the present SN Ia data permits large anisotropy. Recently DE models with variable EOS parameter in anisotropic 
space-time have been studied by Yadav and Yadav \cite{yadav2011}, Yadav et al \cite{yadavetal2011,yadavetal2012}, 
Akarsu and Kilinc 
\cite{akarsu2010}, Yadav \cite{yadav2012}, Saha and Yadav \cite{saha2012} and Pradhan \cite{pradhan2013}. In the 
present work, however, we present $\Lambda$CDM-type cosmological model in spatially homogeneous and anisotropic
 Heckmann-Schucking space-time. The outline of paper is as follows: in section 2, 
the field equation and it's solution are described. 
Section 3 deals with dust filled universe and Hubble's parameter. Section 4 covers the study of 
observational parameters for the model under consideration. The deceleration parameter (DP) and certain 
physical properties of the universe are presented in section 5 and 6 respectively. Finally conclusions are 
summarized in section 7. 

\section{Field equations}.
We consider a general Heckmann-Schucking metric
\begin{equation}\label{metric}
 ds^{2}= c^{2}dt^{2}- A^{2}dx^{2}-B^{2}dy^{2}-C^{2}dz^{2}
\end{equation}
where A, B and C are functions of time only. we consider energy momentum tensor for a perfect fluid  i.e.
\begin{equation}\label{emt}
T_{ij}=(p+\rho)u_{i}u_{j}-pg_{ij}
\end{equation}
where $g_{ij}u^{i}u^{j}=1$ and $u^{i}$ is the 4-velocity vector.\\

In co-moving co-ordinates
  \begin{equation}
    u^{\alpha}=0,~~~~~~~~~\alpha=1,2, 3.
\end{equation}

  The Einstein field equations are
  \begin{equation}\label{efe}
    R_{ij}-\frac{1}{2}Rg_{ij}+ \Lambda g_{ij}= -\frac{8\pi
    G}{c^{4}}T_{ij}
  \end{equation}
  Choosing co-moving coordinates,the field equations (\ref{efe}) in terms of line element (\ref{metric}) can 
be write down as
  \begin{equation}
    \frac{B_{44}}{B}+\frac{C_{44}}{C}+\frac{B_{4}C_{4}}{BC}=-\frac{8\pi G}{c^{2}} p+\Lambda
    c^{2}
  \end{equation}
   \begin{equation}
    \frac{A_{44}}{A}+\frac{C_{44}}{C}+\frac{A_{4}C_{4}}{AC}=- \frac{8\pi G}{c^{2}}p+\Lambda
    c^{2}
  \end{equation}
  \begin{equation}
   \frac{A_{44}}{A}+\frac{B_{44}}{B}+\frac{A_{4}B_{4}}{AB}=- \frac{8\pi G}{c^{2}}p+\Lambda
   c^{2}
  \end{equation}
   \begin{equation}
     \frac{A_{4}B_{4}}{AB}+\frac{B_{4}C_{4}}{BC}+\frac{C_{4}A_{4}}{AC}= \frac{8\pi G}{c^{2}}\rho+\Lambda
     c^{2}
   \end{equation}
where $A_{4}$, $B_{4}$ and $C_{4}$ stand for time derivatives of A, B, and C respectively.\\ 
The mass-energy conservation equation $T^{ij}_{;j}=0$ gives
   \begin{equation}
     \rho_{4}+(p+\rho)(\frac{A_{4}}{A}+\frac{B_{4}}{B}+\frac{C_{4}}{C})=0
   \end{equation}
Subtracting eqs.(5) from (6),(6) from (7) and (7) from (6), we obtain
   \begin{equation}
    \frac{A_{44}}{A}-
    \frac{B_{44}}{B}+\frac{A_{4}C_{4}}{AC}-\frac{B_{4}C_{4}}{BC} = 0
   \end{equation}
   \begin{equation}
     \frac{B_{44}}{B}-
     \frac{C_{44}}{C}+\frac{A_{4}B_{4}}{AB}-\frac{A_{4}C_{4}}{AC} = 0
\end{equation}
\begin{equation}
     \frac{C_{44}}{C}-
     \frac{A_{44}}{A}+\frac{B_{4}C_{4}}{BC}-\frac{A_{4}B_{4}}{AB} = 0
\end{equation}
Subtracting(12) from (10), we get
\begin{equation}
\frac{B_{44}}{B}+\frac{C_{44}}{C}+\frac{2B_{4}C_{4}}{BC}=2\frac{A_{44}}{A}+\frac{A_{4}C_{4}}{AC}+\frac{A_{4}B_{4}}{AB}
\end{equation}
This equation can be re-written in the following form
\begin{equation}
\biggl(\frac{(BC)_{4}}{BC}\biggr)_{4}+\biggl(\frac{(BC)_{4}}{BC}\biggr)^{2}=2\biggl(\frac{A_{4}}{A}\biggr)_{4}+2\frac{A^{2}_{4}}{A^{2}}+\frac{A_{4}(BC)_{4}}{ABC}
\end{equation}
    Integrating this equation, we get the following first
    integral
\begin{equation}
    \bigl(\frac{(BC)_{4}}{BC}-\frac{2A_{4}}{A}\bigr)ABC= L
\end{equation}
where L is constant of integration.\\
The exact solution of eq. (15), in general, is not possible however one can solve 
eq. (15) explicitly, by choosing $L = 0$ that reveals $A^{2} = BC$. The present day observations 
suggest that the initial anisotropy dissipated out for large value of 
time and the directional scale factors have same values in all direction i.e. $A = B = C$ which is 
easily obtained by putting $L = 0$ in eq. (15). That is why $L = 0$ has physical meaning.\\  
Now we can assume
\begin{equation}
    B=AD
\end{equation}
\begin{equation}
    C=AD^{-1}
\end{equation}
where
\begin{equation}
D=D(t)
\end{equation}
Further integrating equation (11), we get the first integral
\begin{equation}
\frac{D_{4}}{D}=\frac{K}{A^{3}}
\end{equation}
where K is an arbitrary constant of integration.\\
With help of equation(15), eq (9) simplifies as
 \begin{equation}
 \rho_{4}+3\frac{A_{4}}{A}(p+\rho)=0
\end{equation}
Thus the Hubble's parameter in this model is
\begin{equation}
H=u^{i}_{;i}=\frac{1}{3}
(\frac{A_{4}}{A}+\frac{B_{4}}{B}+\frac{C_{4}}{C})=\frac{A_{4}}{A}
\end{equation}
Equations(5)-(8) are
simplified as
\begin{equation}
2\frac{A_{44}}{A}+\frac{A^{2}_{4}}{A^{2}}+\frac{K^{2}}{A^{6}}=-\frac{8\pi
G}{c^{2}}p+\Lambda c^{2}
\end{equation}
Taking $K = 0$, we obtain the Einstein's field equation for spatially homogeneous and isotropic 
flat FRW model as
\begin{equation}
\frac{A^{2}_{4}}{A^{2}}-\frac{K^{2}}{3A^{6}}=\frac{8\pi
G}{3c^{2}}\rho+\frac{\Lambda c^{2}}{3}
\end{equation}
\begin{equation}
    2\frac{A_{44}}{A}+\frac{A^{2}_{4}}{A^{2}}=-\frac{8\pi
G}{c^{2}}p+\Lambda c^{2}
\end{equation}
\begin{equation}
\frac{A^{2}_{4}}{A^{2}}=\frac{8\pi G}{3c^{2}}\rho+\frac{\Lambda
c^{2}}{3}
\end{equation}
where A is expansion scale factor.\\
Thus equations(21)-(23) may be regarded as counterpart of FRW
Equations in our anisotropic model. Equations(22) and (23) may be
re-written as
\begin{equation}
2\frac{A_{44}}{A}+\frac{A^{2}_{4}}{A^{2}}=-\frac{8\pi
G}{3c^{2}}\Bigl(p-\frac{\Lambda c^{4}}{8\pi
G}+\frac{K^{2}c^{2}}{8\pi GA^{6}}\Bigr)
\end{equation}
\begin{equation}
H^{2}=\frac{A^{2}_{4}}{A^{2}}=\frac{8\pi
G}{3c^{2}}\Bigl(\rho+\frac{\Lambda c^{4}}{8\pi
G}+\frac{K^{2}c^{2}}{8\pi GA^{6}}\Bigr)
\end{equation}
We now assume that the cosmological constant $\Lambda$ and the
term due to anisotropy also act like energies with densities and
pressures as
$$\rho_{\Lambda}=\frac{\Lambda c^{4}}{8\pi G}\hspace{.5in}
\rho_{\sigma}=\frac{K^{2}c^{2}}{8\pi GA^{6}}$$
\begin{equation}
p_{\Lambda}=-\frac{\Lambda c^{4}}{8\pi G}\hspace{.5in}
p_{\sigma}=\frac{K^{2}c^{2}}{8\pi GA^{6}}
\end{equation}
 It can be easily verified that energy conservation law (20) holds
 separately for $\rho_{\Lambda}$ and $\rho_\sigma$ i.e.
 $$(\rho_{\Lambda})_{4}+3H(p_\Lambda+\rho_\Lambda)=0
 $$\\
$$(\rho_{\sigma})_{4}+3H(p_\sigma+\rho_\sigma)=0
 $$\\
 The equations of state for matter, $ \sigma$ and $\Lambda$ energies are
 as follows
 \begin{equation}
 p_m=\omega_m\rho_m
\end{equation}
where $\omega_m = 0$ for matter in form of dust,
$\omega_m=\frac{1}{3}$ for matter in form of radiation. There are
certain more values of $\omega_m$ for matter in different forms
during the course of evolution of the universe.
 \begin{equation}
    p_{\Lambda}=\omega_{\Lambda}\rho_{\Lambda}
\end{equation}
 Since
 $$p_{\Lambda}+ \rho_{\Lambda}=0$$
  Therefore
  $$\omega_{\Lambda}=-1$$
  Similarly
$$p_{\sigma}=\rho_{\sigma}$$
So,
$$\omega_{\sigma}=1$$
Now we use the following relation between scale factor A and red
shift z
\begin{equation}
    \frac{A_{0}}{A}=1+z
\end{equation}
The suffix(0) is meant for the value at present time.\\
The energy
density $\rho$ comprises of following components
\begin{equation}\label{E.D.}
    \rho=\bigl(\rho_m+\rho_\Lambda+\rho_\sigma \bigr)
\end{equation}
Equations (20) and (31) yield
\begin{equation}
\rho=\sum_i(\rho_i)_0(1+z)^{3(1+\omega_i)}
\end{equation}
Equations (26) and (27) take the form
 \begin{equation}
2\frac{A_{44}}{A}+H^{2}=-\frac{8\pi G}{c^{2}}\Bigl(p_m+p_\Lambda
+p_\sigma\Bigr)
\end{equation}
\begin{equation}
H^{2}=\frac{8\pi
G}{3c^{2}}\bigl(\rho_m+\rho_{\Lambda}+\rho_{\sigma}\bigr)
\end{equation}
\begin{figure*}[thbp]
\begin{tabular}{rl}
\includegraphics[width=8cm]{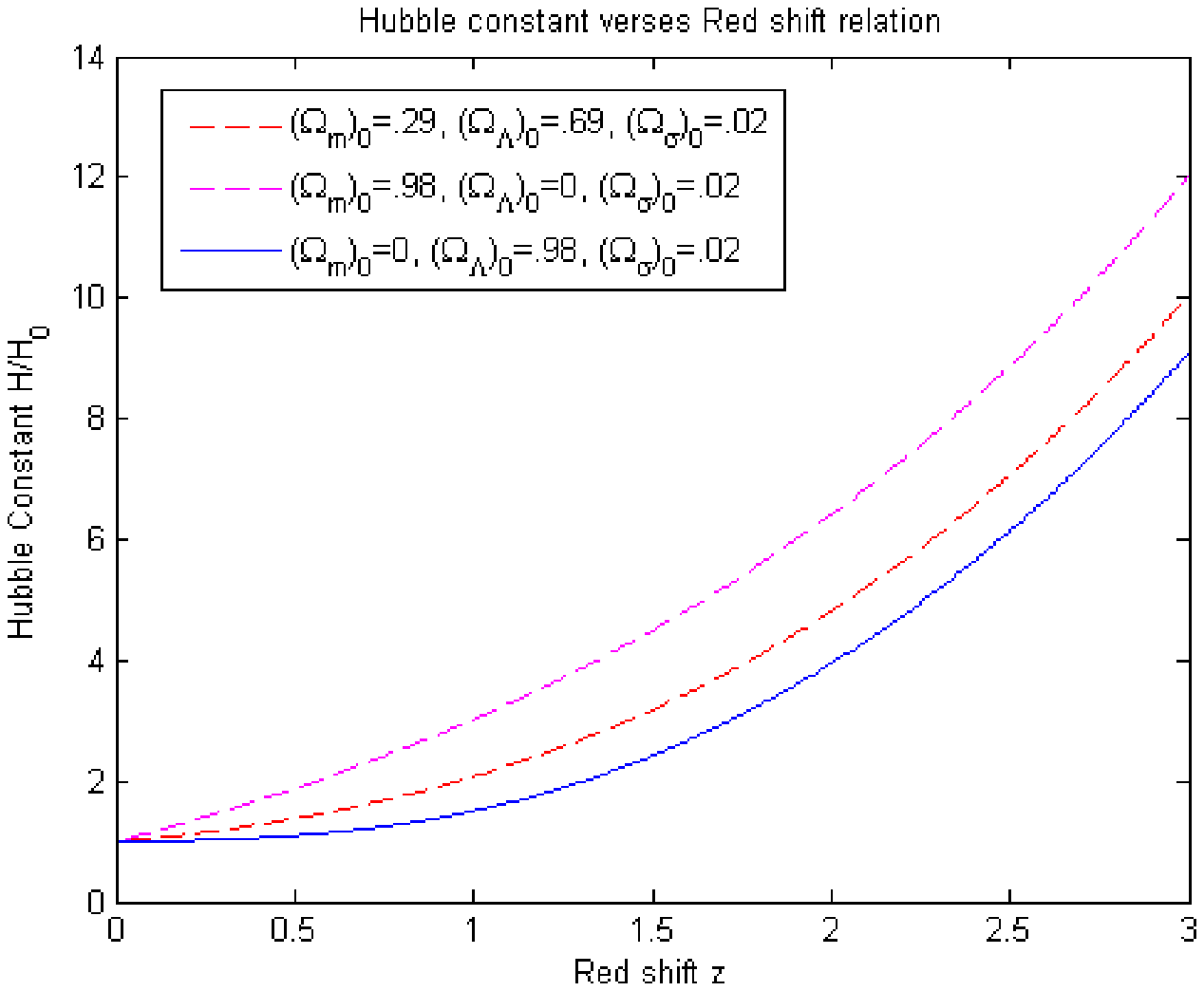}&
\includegraphics[width=8cm]{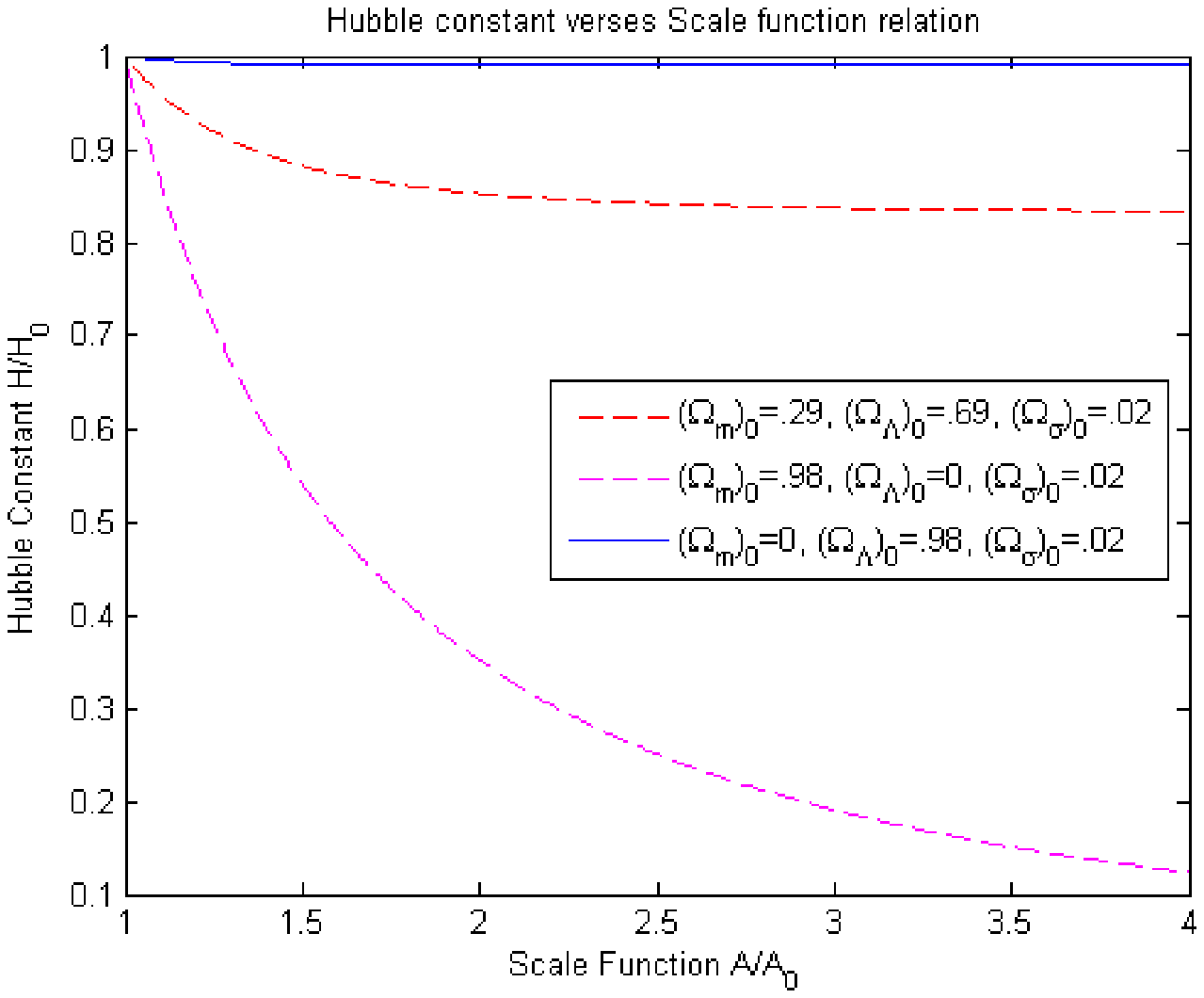}\\
\end{tabular}
\caption{Hubble constant vs redshift $\&$ scale function. }
\end{figure*}
\section{Dust filled universe}
For dust filled universe, we have $p_{m} = 0$ and $\rho_{m} = \frac{3c^{2}H^{2}}{8\pi G}$\\
Then eq. (35) gives
 \begin{equation}
    \Omega_{m}+ \Omega_{\Lambda}+ \Omega_{\sigma}=1
\end{equation}
where $
 \Omega_{m}=\frac{\rho_{m}}{\rho_{c}}
=\frac{(\Omega_m)_0H^2_0(1+z)^3}{H^2}$, $
\Omega_{\Lambda}=\frac{\rho_{\Lambda}}{\rho_{c}}
=\frac{(\Omega_\Lambda)_0H^2_0}{H^2}$ and 
$\Omega_{\sigma}=\frac{\rho_{\sigma}}{\rho_{c}}
=\frac{(\Omega_\sigma)_0H^2_0(1+z)^6}{H^2}$
\subsection{Expression for Hubble's Constant}
Equations (31) and (35) yields
\begin{equation}
H^2 = H_{0}^{2}\left[(\Omega_{m})_{0}(1+z)^{3}+(\Omega_{\sigma})_{0}(1+z)^{6}+(\Omega_{\Lambda})_{0}\right]=
H^{2}_0\bigl[(\Omega_m)_0(\frac{A_0}{A})^3+(\Omega_\sigma)_0(\frac{A_0}{A})^6+(\Omega_\Lambda)_0\bigr]
\end{equation}
The behaviour of Hubble's parameter versus redshift and scale function have been depicted in Figure 1. 
We notice from figure 1 that $\frac{H}{H_{0}}$ increases with redshift while it decreases 
with scale function. From the right panel of figure (1), it is also clear that for $\Lambda$ dominated 
universe, the Hubble's parameter is either almost stationary or it is decreasing slowly with 
scale function i.e. time.  
\section{Some observational constraints}
The luminosity distance which determines flux of
the source is given by
\begin{equation}
    D_{L}=A_{0} x (1+z)
\end{equation}
where $x$ is the spatial co-ordinate distance of a source.\\
The Geodesic for metric (1) ensures that if in the beginning
$\frac{dy}{ds}=0$; $\frac{dz}{ds}=0$ then
$\frac{d^2y}{ds^2}=0$; $\frac{d^2z}{ds^2}=0$.\\
So if a particle moves along x- direction, it continues to move
along x- direction always.If we assume that line of sight of a
vantage galaxy from us is along x-direction then path of photons
traveling through it satisfies
\begin{equation}
ds^{2}= c^{2}dt^{2}- A^{2}dx^2=0
\end{equation}
From this we obtain
\begin{eqnarray*}
  x&=&\int^x_{0}dx=\int^{t_{0}}_{t}\frac{dt}{A(t)}=\frac{1}{A_{0}H_{0}}\int^z_0\frac{dz}{h(z)}  \\
  &=&\frac{1}{A_{0}H_{0}}\int^z_0\frac{dz}{\sqrt{\bigl[(\Omega_m)_0(1+z)^3+(\Omega_\sigma)_0(1+z)^6+(\Omega_\Lambda)_0\bigr]}}
\end{eqnarray*}
\begin{equation}
\end{equation}
where we have used $ dt=dz/\dot{z}$ and from eqs. (21) and (31)
$$\dot{z}=-H(1+z)$$ 
$$ h(z)=\frac{H}{H_0}$$\\
So, the luminosity distance is given by
\begin{equation}\label{LD}
D_{L}=\frac{c(1+z)}{H_{0}}\int^z_0\frac{dz}{\sqrt{\bigl[(\Omega_m)_0(1+z)^3+(\Omega_\sigma)_0(1+z)^6+(\Omega_\Lambda)_0\bigr]}}
\end{equation}

\subsection{Apparent Magnitude and Red Shift relation:} 
The absolute magnitude $(M)$ and apparent magnitude $(m)$ are 
related to the redshift by following relation
\begin{equation}
    m-M = 5log_{10}\bigl(\frac{D_L}{Mpc}\bigr)+25
\end{equation}
For low redshift, one can easily obtain the luminosity distance from eq. (41) 
\begin{equation}
D_L=\frac{cz}{H_0}
\end{equation}
Combining equations (41), (42) and (43), one can easily obtain the expression for the 
apparent magnitude in terms of redshift parameter $(z)$ as follows

\begin{equation}
 m = M+5log_{10}\left(\left(\frac{c(1+z)}{H_{0}Mpc}
\right)\int^{z}_{0}\frac{dz}{\sqrt{[(\Omega_m)_0(1+z)^3+(\Omega_\sigma)_0(1+z)^6+(\Omega_\Lambda)_0]}}\right)
\end{equation}

\begin{figure*}[thbp]
\begin{tabular}{rl}
\includegraphics[width=8cm]{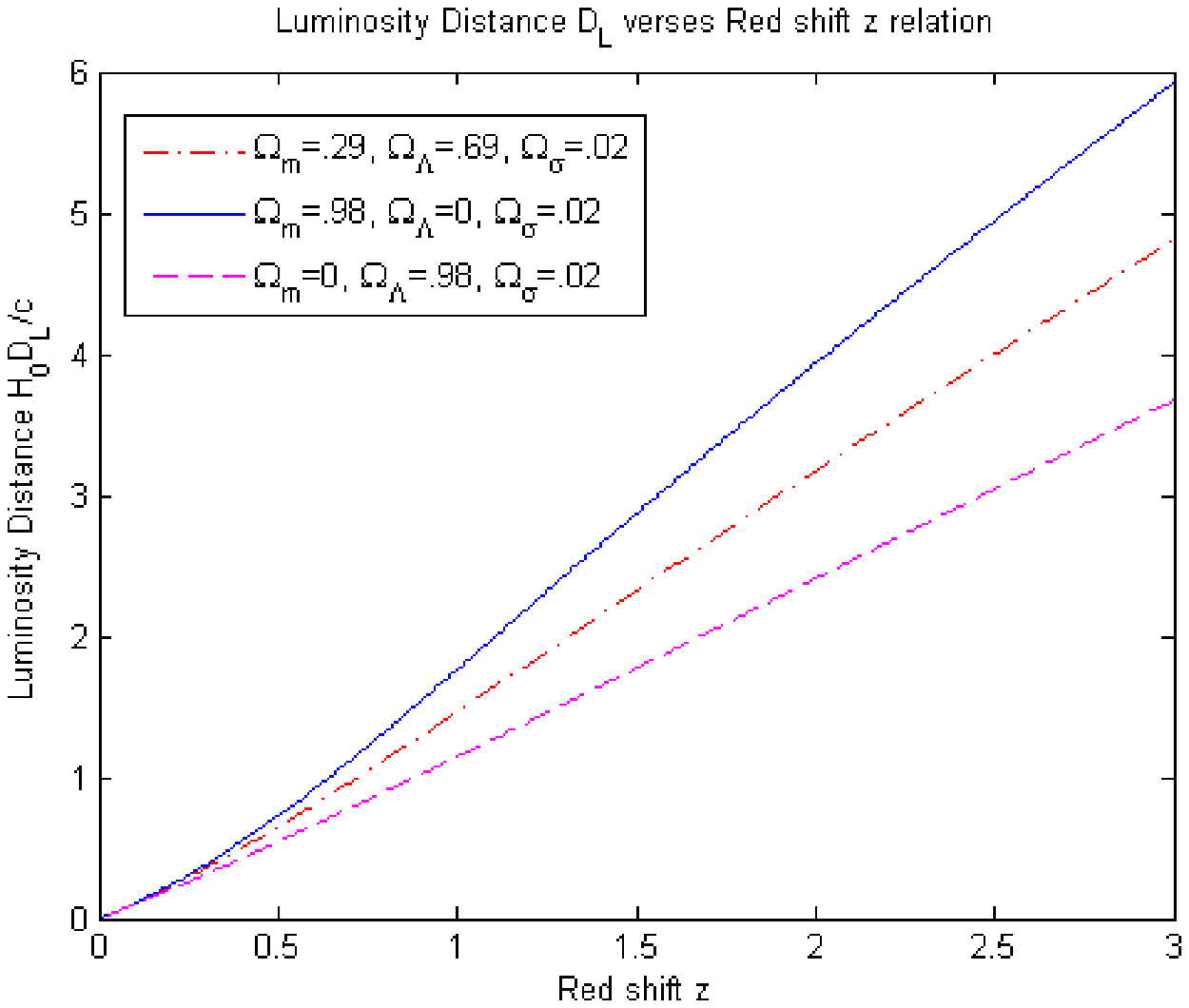}&
\includegraphics[width=8cm]{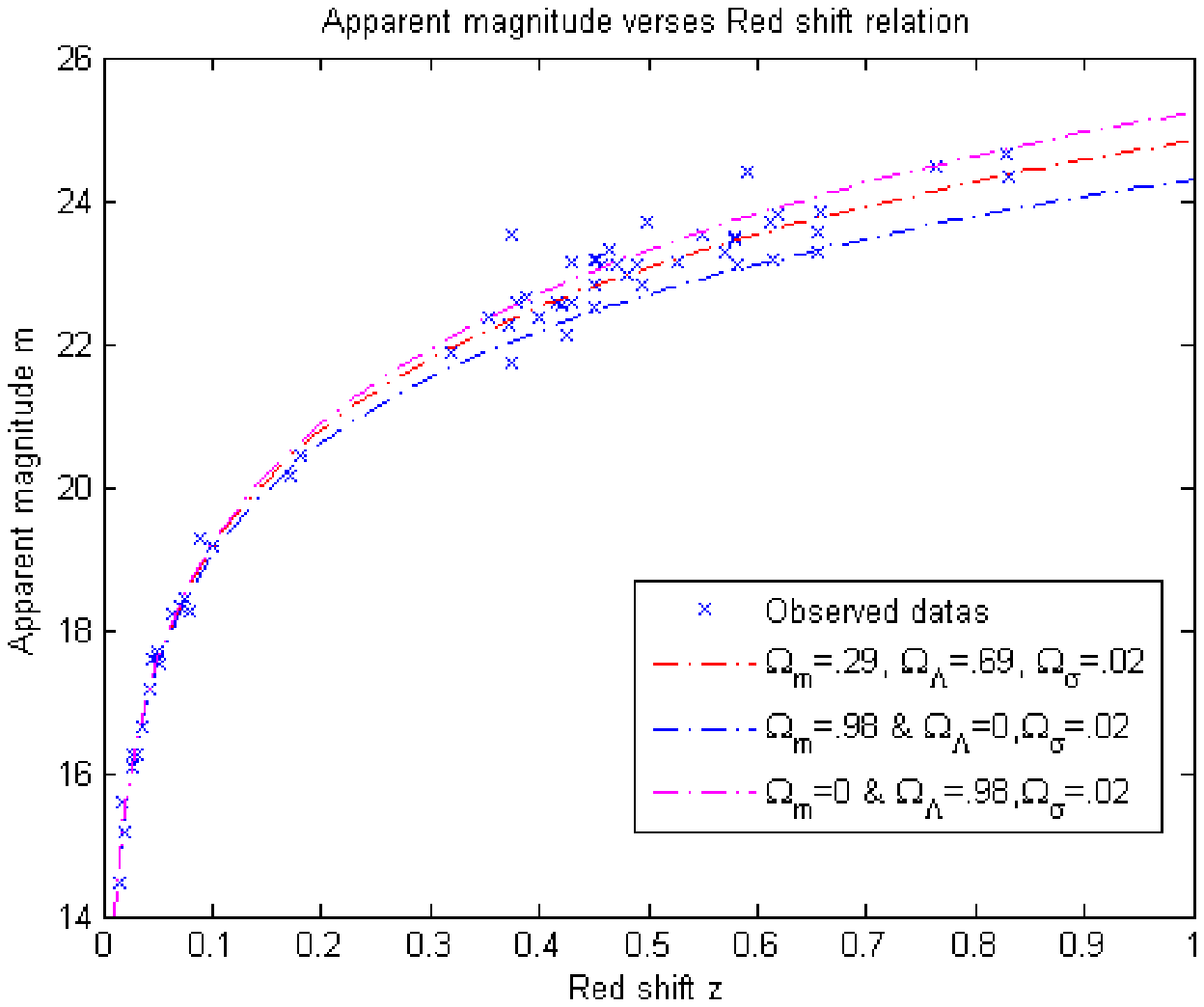}\\
\end{tabular}
\caption{Luminosity distance $\&$ Apparent magnitude vs redshift.}
\end{figure*}
Figure 2 shows the behaviour of luminosity distance $(D_{L})$ and apparent magnitude $(m)$ 
with redshift for some certain values of $\Omega_{m}$, $\Omega_{\Lambda}$ and $\Omega_{\sigma}$.\\

In the present analysis, we use 60 data set of SN Ia for the low red shift $(z < 0.5)$ as reported by 
Perlmutter et al. \cite{Perlmutter1999}. In this case $\chi^{2}_{SN}$ has been computed according to the 
following relation
\begin{center}
{\large{$\chi_{SN}^{2}=A-\frac{B^{2}}{C}+log_{10}(\frac{C}{2\pi})$}}
\par\end{center}{\large \par}

{\large{$ $}}{\large \par}

\begin{center}
{\large{$A=\overset{60}{\underset{i=1}{\sum}}\frac{\left[\left (D_L\right)_{ob}-\left(D_L\right)_{th}\right]^{2}}{\sigma_{i}^{2}}$}}
\par\end{center}{\large \par}

\begin{center}
{\large{$B=\overset{60}{\underset{i=1}{\sum}}\frac{\left[\left(D_L\right)_{ob}-\left(D_L\right)_{th}\right]}{\sigma_{i}^{2}}$}}
\par\end{center}{\large \par}

\begin{center}
{\large{$C=\overset{60}{\underset{i=1}{\sum}}\frac{1}{\sigma_{i}^{2}}$}}
\par\end{center}{\large \par}

$\vphantom{}$
\begin{center}
\begin{tabular}{|c|c|c|c|c|c}
\hline
$(\Omega_{m)_0}$ & $(\Omega_{\Lambda})_0$ & $( \Omega_\sigma)_0$ & $ \chi_{SN}^{2}$ & $\chi_{SN}^{2}/dof$\tabularnewline
\hline
\hline
.29 & .69 & .02 & 7.41417 & 0.1257 \tabularnewline
\hline
.98 & 0 &  .02 &  8.9135  &   0.1511 \tabularnewline
\hline
0 & .98& .02 &  7.8437 &    0.1329 \tabularnewline
\hline
\end{tabular}
\par\end{center}
\begin{center}
\textbf{Table: 1}
\end{center}
Here, dof stands for degree of freedom. From the table 1, we note that the best 
fit values of $(\Omega_{\Lambda})_{0} = 0.69$ with $\chi^{2}_{SN} = 7.41417$ and the reduced $\chi^{2}$ value is 0.1257.  
\subsection{Age of the Universe}
The present age of the universe is obtained as follows
\begin{equation}
   t_{0}=\int^{t_0}_{0}dt=
\int^{\infty}_0\frac{dz}{H_0 (1+z)\sqrt{\bigl[(\Omega_m)_0(1+z)^3+(\Omega_\sigma)_0(1+z)^6+(\Omega_\Lambda)_0\bigr]}} 
\end{equation}
where we have used $ dt=dz/\dot{z}$ and  $\dot{z}=-H(1+z)$\\
The left panel of figure 3 shows the variation of time with redshift. It is also observed that the  
$\Lambda$ dominated universe gives the age of the universe as  $H_ot_0 \simeq .9$\\
Since, $ H_{0}^{-1}=1.3574e+010$\\
$\Rightarrow t_{0}= 0.9(1.357e+010)~Gyrs = 12.30891~Gyrs$.\\
From WMAP data, the empirical value of present age of universe is $13.73_{-.17}^{+.13}$ which is very close 
to present age of universe, estimated in the derived model.
\section{Deceleration Parameter}
The deceleration parameter is given by
\begin{equation}
    q = -\frac{A_{44}}{AH^2}=-\frac{AA_{44}}{A_{4}^{2}}
\end{equation}
 From equation (34)
\begin{eqnarray*}
   -2q + 1&=& -3\sum_{i}\omega_{i}\Omega_{i} \\
   -2q&=& 3\Omega_{\Lambda}-3\Omega_{\sigma}-1
\end{eqnarray*}
This equation clearly shows that without presence of $\Lambda$ term in the Einstein's field 
equation (\ref{efe}), one can't imagine of accelerating universe.This equation also expresses the fact that anisotropy
   raises the lower limit value of $\Lambda$ required for acceleration. This
   may be seen in the following way.\\
   For FRW model,  acceleration requires\\
   \begin{equation}
   \Omega_{\Lambda}\geq.33
\end{equation}
where as for anisotropic model
\begin{equation}
\Omega_{\Lambda}\geq.33+\Omega_{\sigma}
\end{equation}
Combining equations (34), (35), (36) and (37), the expression for DP in terms of redshift $(z)$ is given by
\begin{equation}
q = \frac{3}{2}\Bigl( \frac{(\Omega_m)_o (1+z)^3 +
2(\Omega_\sigma)_o (1+z)^6}{(\Omega_m)_o
(1+z)^3+(\Omega_\Lambda)_o + (\Omega_\sigma)_o
 (1+z)^6}\Bigr)-1
\end{equation}
Since in the derived model, the best fit values of $(\Omega_{m})_{0}$, $(\Omega_{\Lambda})_{0}$ 
and  $(\Omega_{\sigma})_{0}$ are 0.29, 0.69 and 0.02 respectively hence we compute 
the present value of DP for derived $\Lambda$CDM universe by putting $z = 0$ in eq. (49). The 
present value of DP is given by
\begin{equation}
 q_{0} = -0.505
\end{equation} 
\section{Some Physical Properties of the Model}
\subsection{The energy density in the universe}
The energy density $\rho$ is given by
\begin{center}
\begin{equation}
\rho=\underset{i}{\sum}(\rho_{i})_{0}(1+z)^{3\left(1+\omega_{i}\right)}
\end{equation}
\par\end{center}
where
\begin{equation}
(\rho_{i})_{0}=\frac{3c^{2}H_{0}^{2}}{8\pi G}(\Omega_{i})_{0}
\end{equation}
Here $(\rho_{i})_{0}$ are the present energy density of various
components.
Taking,
 $$(\Omega_{m})_{0} \simeq .3, (\Omega_{\Lambda})_{0} \simeq .7, H_{0}=72km/sec./Mpc$$ 
Therefore, the present value of dust energy density $(\rho_{m})_{0}$ and dark energy 
density $(\rho_{\Lambda})_{0}$ are obtained as
\begin{equation}
(\rho_{m})_{0}=2.8727\times10^{-19}gm/cm^{3}
\end{equation}
\begin{equation}
(\rho_{\Lambda})_{0}=\frac{(\Omega_{\Lambda})_{0}}{(\Omega_{m})_{0}}(\rho_{m})_{0}=6.7030\times10^{-19}gm/cm^{3}
\end{equation}
\subsection{Shear Scalar}
The shear scalar is given by
\begin{equation}
\sigma^2=\frac{1}{2}\sigma_{ij}\sigma^{ij}
\end{equation}
where
\begin{equation}
    \sigma_{ij}= u_{i;j}-\Theta(g_{ij}-u_iu_j)
\end{equation}
In our model
\begin{equation}
    \sigma^2= \frac{D^2_4}{D^2}=
    \frac{K^2}{A^6}=3(\Omega_\sigma)_0H^2_0(1+z)^6
\end{equation}
From eq. (57), it is clear that shear scalar vanishes as $A\rightarrow\infty$.
\begin{figure*}[thbp]
\begin{tabular}{rl}
\includegraphics[width=8cm]{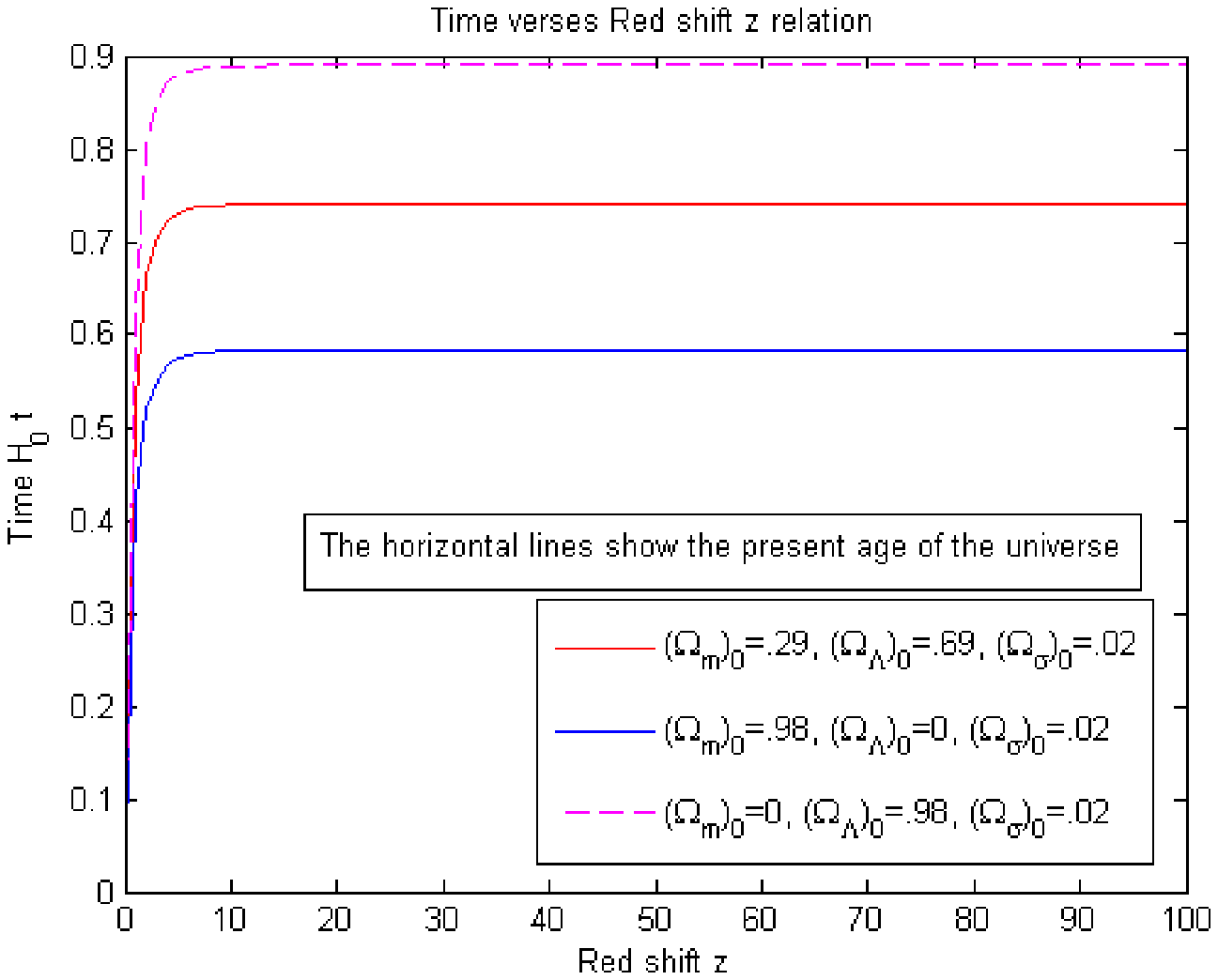}&
\includegraphics[width=8cm]{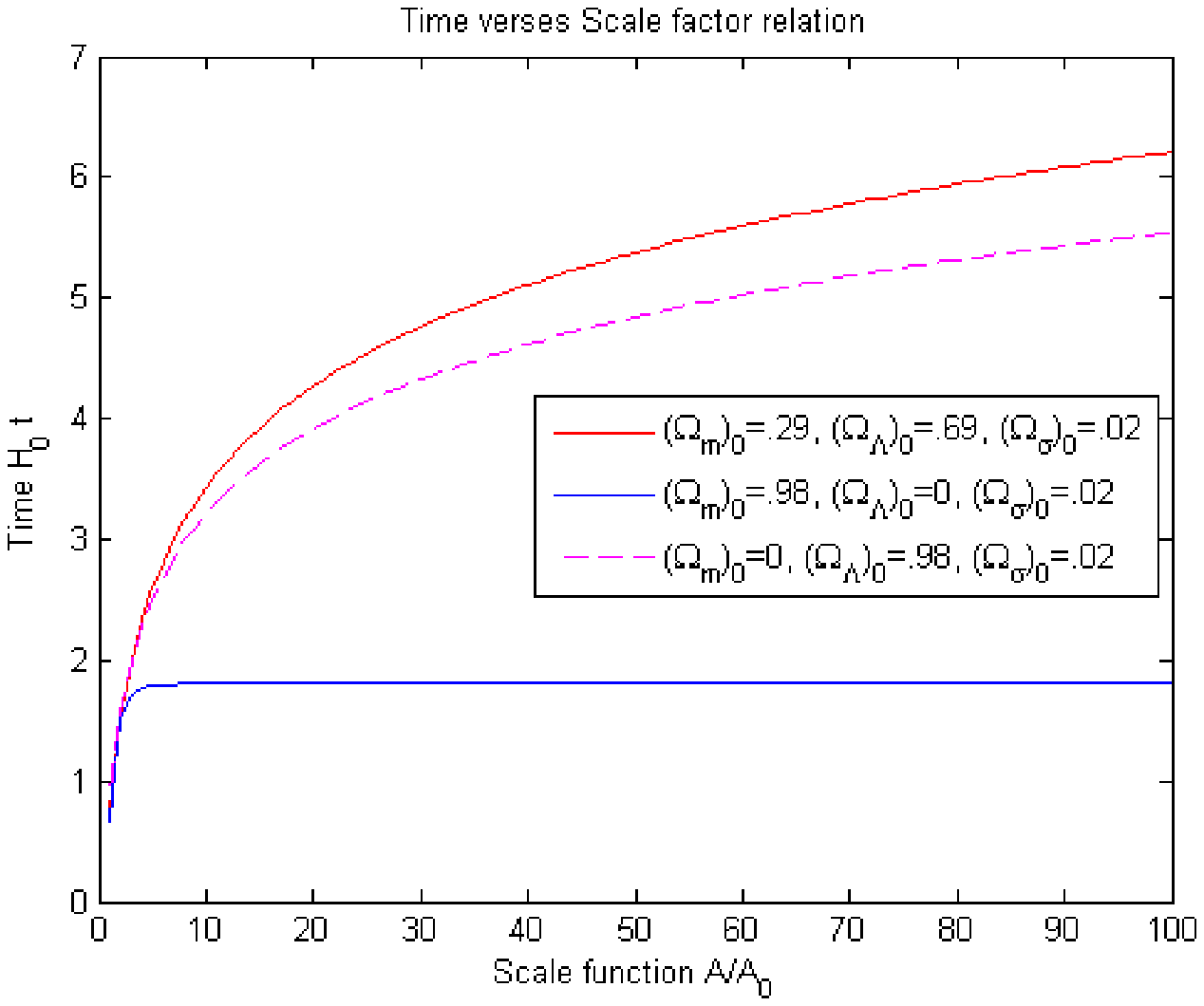}\\
\end{tabular}
\caption{Time vs redshift $\&$ scale function.}
\end{figure*}
\subsection{Relative Anisotropy}
The relative anisotropy is given by
\begin{equation}
    \frac{\sigma^2}{\rho_m}=\frac{3(\Omega_\sigma)_0H^2_0(1+z)^3}{(\rho_c)_0(\Omega_m)_0}
\end{equation}
This follows the same pattern as shear scalar. This means that relative anisotropy
decreases over scale factor i.e. time.\\
 \subsection{Evolution of the scale factor $(A)$}
 We begin with the integral
\begin{equation}
t  =  \intop_{0}^{t}dt 
 = \intop_{0}^{A}\frac{dA}{AH}
\end{equation}
Equations (37) and (59) lead to
\begin{equation}
t=\intop_{0}^{A}\frac{dA}{AH_0\sqrt{\bigl[((\Omega_m)_0(\frac{A_0}{A})^3++(\Omega_\sigma)_0(\frac{A_0}{A})^6
+(\Omega_\Lambda)_0\bigr]}}
\end{equation}
The right panel of figure 3 shows the variation of time with scale function for the derived model.\\
\section{Final remarks}
In this paper, we have investigated the $\Lambda$CDM-type cosmological model in Heckmann-Shucking space-time. 
Under some specific choice of parameters, the model under consideration isotropizes and have consistency with 
recent SN Ia observation. We have estimated some physical parameters at present epoch for derived model which 
is summarized as 
\begin{center}
\begin{tabular}{|c|c|}
\hline $t_{0}$ & 12.30891 Gyrs\\
\hline $q_{0}$ & -0.505\\
\hline $(\rho_{m})_{0}$ & $2.8727 \times 10^{-19}~ gm/cm^{3}$\\
\hline $(\rho_{\Lambda})_{0}$ & $6.7030 \times 10^{-19}~ gm/cm^{3}$\\
\hline 
\end{tabular}
\end{center}
\begin{center}
 \textbf{Table: 2}
\end{center}
We observe from the result displayed in table 2 that the derived model is observationally indistinguishable in 
the vicinity of present epoch of universe. Thus the $\Lambda$CDM model fits better with the observational data.\\

\end{document}